\documentclass[twocolumn,preprintnumbers,superscriptaddress,endnote,nofootinbib,aps,prd,floatfix]{revtex4}

\usepackage{footmisc}
\usepackage{multirow}
\usepackage{subfigure}
\usepackage{amsmath}
\usepackage{enumitem}

\usepackage{graphicx}

\newcommand{\be}{\begin{eqnarray*}}
\newcommand{\ee}{\end{eqnarray*}}
\newcommand{\gl}[1]{(\ref{#1})}

\newcommand{\bee}{\begin{eqnarray}}
\newcommand{\eee}{\end{eqnarray}}
\newcommand{\beeq}{\begin{equation}}
\newcommand{\eeeq}{\end{equation}}

\newenvironment{cedescription}{%
   
   \begin{description}[leftmargin=0.25cm, style=sameline]%
}{%
   \end{description}%
}

\newenvironment{mydescription}{%
   
   \begin{description}[leftmargin=0.25cm, style=sameline]%
}{%
   \end{description}%
}

\preprint{IPPP/14/77} \preprint{DCPT/14/154}

\begin{document}

\title{Effective Theories and Measurements at Colliders}

\begin{abstract}
  If the LHC run 2 will not provide conclusive hints for new resonant
  Physics beyond the Standard Model, dedicated and consistent search
  strategies at high momentum transfers will become the focus of
  searches for anticipated deviations from the Standard Model
  expectation.  We discuss the phenomenological importance of QCD and
  electroweak corrections in bounding higher dimensional operators
  when analysing energy-dependent differential distributions. In
  particular, we study the impact of RGE-induced operator running and
  mixing effects on measurements performed in the context of an
  Effective Field Theory extension of the SM. Furthermore we outline
  a general analysis strategy which allows a RGE-improved formulation
  of constraints free of theoretical shortcomings that can arise when
  differential distributions start to probe the new interaction
  scale. We compare the numerical importance of such a programme
  against the standard analysis approach which is widely pursued at
  present.
\end{abstract}

\author{Christoph Englert} \email{christoph.englert@glasgow.ac.uk}
\affiliation{SUPA, School of Physics and Astronomy,University of
  Glasgow,\\Glasgow, G12 8QQ, United Kingdom\\[0.1cm]}

\author{Michael Spannowsky} \email{michael.spannowsky@durham.ac.uk}
\affiliation{Institute for Particle Physics Phenomenology, Department
  of Physics,\\Durham University, DH1 3LE, United Kingdom\\[0.1cm]}

\maketitle


\section{Introduction}
\label{sec:intro}
After the Higgs discovery in 2012~\cite{hatlas,hcms}, the ATLAS and
CMS collaborations have started to investigate the new particle's
properties in further detail~\cite{couplings_ex}. For run 1, the Higgs
boson's couplings have been constrained primarily using ratios
\begin{equation}
  \label{eq:kappa}
  \kappa = (g_{\text{SM}}+\Delta g_{\text{BSM}})\,
  /g_{\text{SM}}
\end{equation}
see Ref.~\cite{LHCHiggsCrossSectionWorkingGroup:2012nn} for
details. These quantities are inclusive with respect to the phase
space and are determined by comparing the number of measured events
with the Standard Model prediction after subtracting the background
for a given process. While this strategy is a reasonable procedure to
obtain limits with relatively small statistics and large systematic
uncertainties, a larger parameter space will become accessible during
run 2, and a more fine-grained picture of constraints on interactions
beyond the SM (BSM) can be formulated at higher LHC luminosity and
energy.

In the absence of new resonant effects, a common approach to
parametrise new physics interactions is to employ effective theory
methods~\cite{dim6,dim6e,dim6r,dim6g}. Imposing simplifying
assumptions, such as {\it e.g.} the absence of non-trivial BSM flavour
structures, one obtains a basis of 59 independent operators that
express our lack of knowledge of the underlying new physics model at a
high scale~\cite{dim6r}.

New physics at energy scales larger than the electroweak scale will
typically show up as modifications of differential distributions at
high transverse momenta. While an increased cross section can be
observable in inclusive ``$\sigma\times \text{BR}$ physics'', a
proper investigation of differential distributions is not only far
more adequate to this particular physics question, but will also
provide significantly more insight into the nature of BSM physics if a
significant excess over the SM will be observed eventually.

A clear advantage of abandoning the $\kappa$ prescription of
Eq.~\eqref{eq:kappa} in favour of an effective field theory approach
with a general set of Higgs interaction operators is that information
from differential distributions does have a theoretically meaningful
interpretation. The presence of dimension 6 operators will not only
alter the total rate, but also the shape of measured distributions and
new physics searches (in the absence of new kinematically accessible
resonances) can be studied in a fairly model independent way.

However, there are a few caveats. Using differential distributions can
also mean a challenge for the effective theory approach. Effective
theory, being an expansion in a new physics interaction scale
$\Lambda_{\text{NP}}$, is strictly speaking only valid when typical
interaction scales are distinctively separated, {\it i.e.} when we
have $\Lambda_{\text{NP}} \gg \Lambda_{\text{interaction}}$ for all
relevant scales of the considered process. A well-known example for
this is flavour physics, where effective field theories have always
been an important tool. When studying rare decays, the weak
interaction scale $\Lambda_{\text{NP}}=m_W$ is clearly separated from
the scale at which B-Mesons decay $\Lambda_{\text{interaction}}=m_b$,
which acts as the characteristic measurement scale. Corrections to the
effective theory description are parametrically suppressed by
$\mathcal{O}(m_b^2/m_W^2) \sim 0.3\%$. Therefore, applying effective
field theory methods provides a well-motivated and theoretically
well-controlled approximation.

At collider experiments in general, but at hadron colliders in
particular, it is challenging to infer the scale at which the
effective operators are probed from the observed final state objects
when we want to formulate a limit on the presence of new physics:
Different events will always probe the theory prediction at different
scales $\mu$. For example, in mono-Higgs production where the Higgs
recoils against a hard jet, the transverse momentum of the jet is a
relevant scale at which the effective operator $ \hat H^\dagger \hat H
\hat{G}^{\mu\nu}\hat{G}_{\mu \nu}/\Lambda_{\text{NP}}^2$ is probed.

On the one hand, a naive constraint on $C_O
(v^2/\Lambda_{\text{NP}}^2)$ can always be understood as a limit
obtained with $\Lambda\ll \Lambda_{\text{NP}}$ with an appropriate
redefinition of the Wilson coefficient's size and we even might be
tempted to lower $\Lambda_{\text{NP}}$ to an energy range of a few TeV
that is resolved by the LHC for an educated guess of the Wilson
coefficient.\footnote{This procedure has typically been applied in
  searches for Dark Matter at the LHC and has been left without
  criticism for quite some time~\cite{Buchmueller:2013dya}.} The
reliability and robustness of such a limit is at least questionable as
a naive analysis of a Wilson coefficient is performed completely
independent of the matching or cut-off scale, which must not be
kinematically resolved for the EFT expansion to hold in the first
place. 

On the other hand, if the effective Lagrangian is defined at a fixed
scale $\Lambda_{\text{NP}}$ {\it{outside}} the LHC reach or the
observable's energy coverage\footnote{This situation is similar to
  electroweak fits after LEP2, which assumed a Higgs mass at the
  kinematic endpoint of $m_H\simeq 114$~GeV.}, or at least at the
maximum energy probed in a new physics experiment with negative
outcome, they mix when evolved from one scale to another as a
consequence of electroweak and QCD
interactions~\cite{trott1,Jenkins:2013zja,espinosa}. As a result,
different phase space regions do probe different operator
combinations. Thus, to infer well-defined constraints from exclusive
distributions, the operators probed at different energy scales for
different events or bins have to be evolved to a fixed predefined
scale to allow a direct interpretation.

The impact of operator running is parametrically $ \mathcal{O} \left
  (g_i \gamma_i \log \left [ \Lambda_{\text{NP}} /
    \Lambda_{\text{meas}} \right] \right )$, with coupling $g_i$ and
the anomalous dimension $\gamma_i$ of the operator $\hat{O}_i$, the new
physics scale $\Lambda_{\text{NP}}$ and the measurement scale
$\Lambda_{\text{meas}}$. For B-decay observables with
$\Lambda_{\text{NP}} \simeq m_W$ and $\Lambda_{\text{meas}} \simeq
m_b$, the resummation of these large logarithms can provide an
important theoretical improvement for the interpretation of the
measurement. A priori, when studying Higgs boson properties and
assuming no New Physics particles up to several TeV, the hierarchy of
electroweak and New Physics scale ({\it e.g.} $\Lambda_{\text{NP}}
\simeq 2~\text{TeV}$) can be of similar order. Hence a resummation of
these large logarithms can be crucial for a detailed understanding of
the impact of Higgs-boson measurements on New Physics models. 

\medskip

In this paper, we study the impact of operator running and mixing on
coupling measurements using differential distributions. We focus on
three illustrative examples ranging from multi-jet to Higgs physics.
To our knowledge these effects have not been discussed in a fully
differential fashion at the LHC in the context of effective field
theory measurements. We also provide a first step towards a general
prescription of how measurements based on differential distributions
can be used to constrain an effective Lagrangian, and how to give
those constraints an interpretation in terms of a UV scale model,
including higher-order corrections in a well-defined and practical
way. As we will see, due to the momentum dependence of many of the
higher-dimensional operators and their impact being most relevant when
probed at large invariant masses, {\it i.e.}
$\Lambda_{\text{meas}}\simeq \sqrt{\hat{s}}$, the characteristic
logarithms $\log ( \Lambda_{\text{NP}} / \sqrt{\hat{s}})$, depending
on the assumed new physics scale $\Lambda_{\text{NP}}$, are fairly
small and the contributions of operator running is of $\lesssim 10
\%$.

To make this work self-contained we review the (flavour physics)
language relevant to this problem in Sec.~\ref{sec:eff} before we
apply it to di-jet final states at the LHC. In
Secs.~\ref{sec:Higgsjet} and \ref{sec:Higgsass} we discuss the impact
on Higgs phenomenology in $H$+jet and $HZ$ production before we give
our conclusions in Sec.~\ref{sec:conc}.


\section{Effective Field Theory Approach: A Quick Review}
\label{sec:eff}
In general an effective Hamiltonian in Operator Product Expansion is
given by
\begin{equation}
  \label{eq:split}
  \hat{\mathcal{H}}_{\text{eff}} = \sum_i C_i (\mu) \hat{O}_i(\mu)\,,
\end{equation}
where ${\hat O}_i$ are the operators defined at the factorisation
scale $\mu$ and $C_i$ are the so-called Wilson coefficients. Note that
as a consequence of factorisation, both the Wilson coefficient as well
as the operators are scale-dependent. This dependence cancels for
$\hat{\mathcal{H}}_\text{eff}$. Eq.~\gl{eq:split} separates the
physics into a long-range behaviour of matrix elements $\langle {\hat
  O}(\mu)\rangle$ and short-range behaviour of Wilson coefficients
$C_i(\mu)$ relative to the factorisation scale $\mu$. The ignorance of
physics with respect to this arbitrary separation at this stage leads
to renormalisation group equations (RGEs). If we focus on a particular
model, the coefficients of Eq.~\gl{eq:split} can be obtained by a
matching calculation. Only assuming SM particle content and gauge
symmetries, the lowest order effective operator extension consists of
dimension 6 operators documented in Ref.~\cite{dim6r}. Relying on this
language, we are fairly unprejudiced about the particular UV dynamics
at a new physics scale $\Lambda_{\text{NP}}$ (a well-motivated guess
on the Wilson coefficients' hierarchies are possible when we consider
composite Higgs scenarios~\cite{dim6g}).

Approximating general amplitudes and eventually exclusive cross
sections in terms of effective operators is only valid if the new
physics scale $\Lambda_{\text{NP}}$, the scale of the masses of the
heavy degrees of freedom of the full theory, is much larger than the
scale at which the effective operator is probed
(see~\cite{hzdim6,hzdim62,riva,gero} for a discussion in the context
of Higgs physics).

For example, in the Standard Model process $c\bar{s} \to u \bar{d}$
the leading-order amplitude is given by (we suppress the CKM matrix
elements for convenience)
\begin{eqnarray}
  \mathcal{M} &=& i \frac{G_F}{\sqrt{2}} \frac{M_W^2}{\hat{s} - M_W^2} 
  (\bar{s}_ac_a)_{V-A}(\bar{u}_bd_b)_{V-A} \nonumber \\ 
  &=& - i \frac{G_F}{\sqrt{2}}
  (\bar{s}_ac_a)_{V-A}(\bar{u}_bd_b)_{V-A} 
  + \mathcal{O} \left ( \frac{\hat{s}}{M_W^2} \right ),
\label{eq:4famp}
\end{eqnarray}
assuming a diagonal CKM matrix and $(V-A)$ referring to the Lorentz
structure $\gamma_\mu (1 - \gamma_5)$ (we have made the color indices
$a$ and $b$ of the spinors explicit). While the expansion using an
effective operator ${\hat O}_2=(\hat{\bar{s}}_a\hat
c_a)_{V-A}(\hat{\bar{u}}_b\hat d_b)_{V-A}$ preserves the Lorentz
structure of the interaction, the kinematics due to the exchange of a
$W$ boson is omitted, which is only valid if the partonic centre of
mass energy $\hat{s}=(p_{\bar{s}}+p_c)^2 \ll M_W^2$.
 
It is worth noting that higher-order corrections can increase the
numbers of operators necessary to describe a process. Including higher
order corrections to the amplitude in Eq.~\gl{eq:4famp}, {\it i.e.}
gluon exchange between different quark legs, two linearly independent
operators, indicating a different color flow, will contribute to the
amplitude. After performing a matching calculation between the
effective and full theory, we can express the amplitude in leading-log
approximation using effective operators as~\cite{Buchalla:1995vs}
\begin{eqnarray}
  i\mathcal{M} &=&  C_1  \langle {\hat O}_1  \rangle 
  + C_2  \langle {\hat O}_2  \rangle
\label{eq:4fnlo}
\end{eqnarray}
with renormalised\footnote{For a discussion of the renormalisation
  scheme dependence of Wilson coefficients
  see~\cite{Buchalla:1995vs}.}  matrix elements $\langle {\hat O}_i
\rangle$ 
\begin{eqnarray}
\frac {\sqrt{2}}{G_F}  \langle {\hat O}_1 \rangle &=& \left ( 1+ 2 C_F \frac{\alpha_s}{4 \pi} \log \frac{\mu^2}{\hat{s}} \right ) S_1 \nonumber \\
&+& 
 \frac{\alpha_s}{4 \pi} \log \frac{\mu^2}{\hat{s}} S_1 - 3 \frac{\alpha_s}{4\pi} \log \frac{\mu^2}{\hat{s}}S_2, \nonumber \\
\frac {\sqrt{2}}{G_F} \langle {\hat O}_2 \rangle &=& \left ( 1+ 2 C_F \frac{\alpha_s}{4 \pi} \log \frac{\mu^2}{\hat{s}} \right ) S_2 \nonumber \\
&+& 
\frac{\alpha_s}{4\pi} \log \frac{\mu^2}{\hat{s}}S_2 - 3 \frac{\alpha_s}{4 \pi} \log \frac{\mu^2}{\hat{s}}S_1,
\label{eq:4fopnlo}
\end{eqnarray}
and 
\begin{eqnarray}
S_1 &=& (\bar{s}_a c_b)_{V-A} (\bar{u}_b d_a)_{V-A}\,, \nonumber \\
S_2 &=& (\bar{s}_a c_a)_{V-A} (\bar{u}_b d_b)_{V-A}\,,
\end{eqnarray}
where $C_F,C_A$ are the casimirs of the fundamental and adjoint
representations respectively. The Wilson coefficients $C_1$ and $C_2$,
as a result of the matching calculation, are given
by~\cite{Buchalla:1995vs}
\begin{eqnarray}
C_1 &=& -3 \frac{\alpha_s}{4 \pi}\log \frac{M_W^2}{\mu^2} \nonumber \\
C_2 &=&1 + 
\frac{\alpha_s}{4\pi} \log \frac{M_W^2}{\mu^2}.
\label{Eq:WilsonCoeff1}
\end{eqnarray}

As the operators in this process have mass dimension 6, the couplings
have the form $C_i \langle {\hat O}_i \rangle / \Lambda_{\text{NP}}^2
\sim g^2_i/M^2_W$, {\it i.e.} they represent ratios between
dimensionless couplings and the validity scale of the effective
theory.  The unphysical factorisation scale $\mu$ in
Eqs.~(\ref{eq:4fopnlo}) and (\ref{Eq:WilsonCoeff1}) constitutes a
formal separation between long and short distance physics, {\it i.e.}
the matrix element $\langle {\hat O}_i \rangle$ and the Wilson
coefficient $C_i$. It becomes obvious that, depending on the ratio of
the two scales $\hat{s}$ and $M^2_W$ and the choice of $\mu$, large
logarithms of the form $\alpha_s(\mu)\log(M_W^2/\mu^2)$ can appear
which degrade the reliability of the fixed-order calculation.

To enhance the reliability of the perturbative series one typically
reverts to RG-improved calculations that partially resum these
logarithms to a given formal logarithmic and perturbative
accuracy. This yields an improved formulation of physics at energies
significantly lower than the new physics scale $\Lambda = M_W$, for
example the scale $\sqrt{\hat s}$ at which the measurement is performed.

\medskip

As we perform a calculation in EFT with higher dimensional bare
interactions $\sim C_i^{(0)}{\hat O}_i(\hat{\bar u}^{(0)}\hat{d}^{(0)}
\hat{\bar s}^{(0)}\hat{c}^{(0)})$ there is an additional
multiplicative renormalisation of the Wilson coefficients necessary to
arrive at the above result; just like all couplings of the
renormalisable part of the Lagrangian we can think of the Wilson
coefficients as coupling constants that can be renormalised in a
straightforward fashion (we will discuss explicit examples further
below).\footnote{The anomalous dimension for {\it e.g.} the strong
  coupling $g_s$ is the $\beta$ function divided by $g_s$ due to a
  choice of conventions.} This renormalisation implies the mentioned
RGE for the Wilson coefficients
\begin{equation}
  \label{eq:eqrunn}
  \frac{\text{d}C_i}{\text{d} \log \mu} =  \gamma_{ij} C_j\,,
\end{equation}
where the anomalous dimension matrix $\gamma_{ij}$ is typically
non-diagonal, hence leading to scale-dependent operator mixing under
the RG flow. The RGEs resum potentially large logarithms that arise
from evolving the operator from where we define the EFT to the scale
at which we probe it. Thus, the evolution is entirely encoded in
$\gamma_{ij}$ and the Wilson coefficients at different
    scales $\sqrt{\hat s}$, following from the solution of
    Eq.~\eqref{eq:eqrunn} is given to leading log approximation by
\begin{equation}
  C_i(\sqrt{\hat s}) \simeq \left(\delta_{ij} + 
    \gamma_{ij}(\sqrt{\hat s})
    \log { \sqrt{\hat s} \over \mu } \right)  C_j(\mu)\,.
\end{equation}

The anomalous dimension is related to the
multiplicative (counter term) renormalisation of the bare couplings
$C^{(0)}= Z_C C = (1+\delta_{C}) C$.
\begin{equation}
  \gamma = - \lim_{\varepsilon \to 0}\frac{\text{d} \log Z_C} {\text{d}\log \mu} 
\end{equation}
in dimensional regularization with $D=4-2\varepsilon$. At the one loop
level we can replace $Z_C$ by the counter term $\delta_C$ for the
Wilson coefficient. 

For the discussed case of Eqs~(\ref{eq:4famp})-(\ref{Eq:WilsonCoeff1})
the anomalous dimension matrix reads
\begin{equation}
  \label{eq:4fermgam}
  \gamma={1 \over 16\pi^2} \left[
    \begin{matrix}
      -2g_s^2 & 6g_s^2 \\
      6g_s^2 & -2g_s^2 \\
    \end{matrix}
  \right]\,.
\end{equation}
It can be diagonalised in a straightforward fashion, yielding a
decoupled set of RGEs for the linear combination of Wilson
coefficients $C_\pm= C_2\pm C_1$.


\section{Constraining New Physics by Measuring Wilson Coefficients}
\label{sec:wilson}
In flavour physics, where this conceptual apparatus has been put to
good use for the last decades~\cite{flavorpapers}, the lower
characteristic scale usually corresponds to the mass of the decaying
quark of the hadron whose properties are to be studied, {\it e.g.}
$\hat{s} = m^2_b$. In contrast to that, at the LHC fixing the lower
(IR) scale, {\it e.g.} $\hat{s} = m^2_H$, is not possible in all
analyses. The range of $\hat{s}$ probed at the LHC, even for a single
observable, can be large and extend easily to the multi-TeV regime as
encountered in {\it e.g.} Higgs+jet
phenomenology~\cite{uli,schlaffer_spannowsky}. Therefore, due to
operator running and mixing, each event probes a different combination
of operators at $\hat{s}$. These measurements or constraints have to
be related to the operators defined at the new physics scale
$\Lambda_{\text{NP}}$ to allow a consistent formulation of a combined
constraint on a new physics model defined at this scale.

Being able to constrain or discover new physics contributions in
differential distributions, {\it i.e.} measurements beyond total
rates, is a particularly intriguing feature of the LHC with its large
centre of mass energy and its increasing amount of integrated
luminosity. During the upcoming runs at $13$-$14$ TeV we can expect the
focus of BSM searches to quickly move towards constraining EFTs with
the help of differential distributions.

\begin{figure}[!t]
  \centering
    \parbox{0.48\textwidth}{     
      \vspace{-0.8cm}
      \hspace{-0.4cm}
      \includegraphics[angle=270,width=0.49\textwidth]{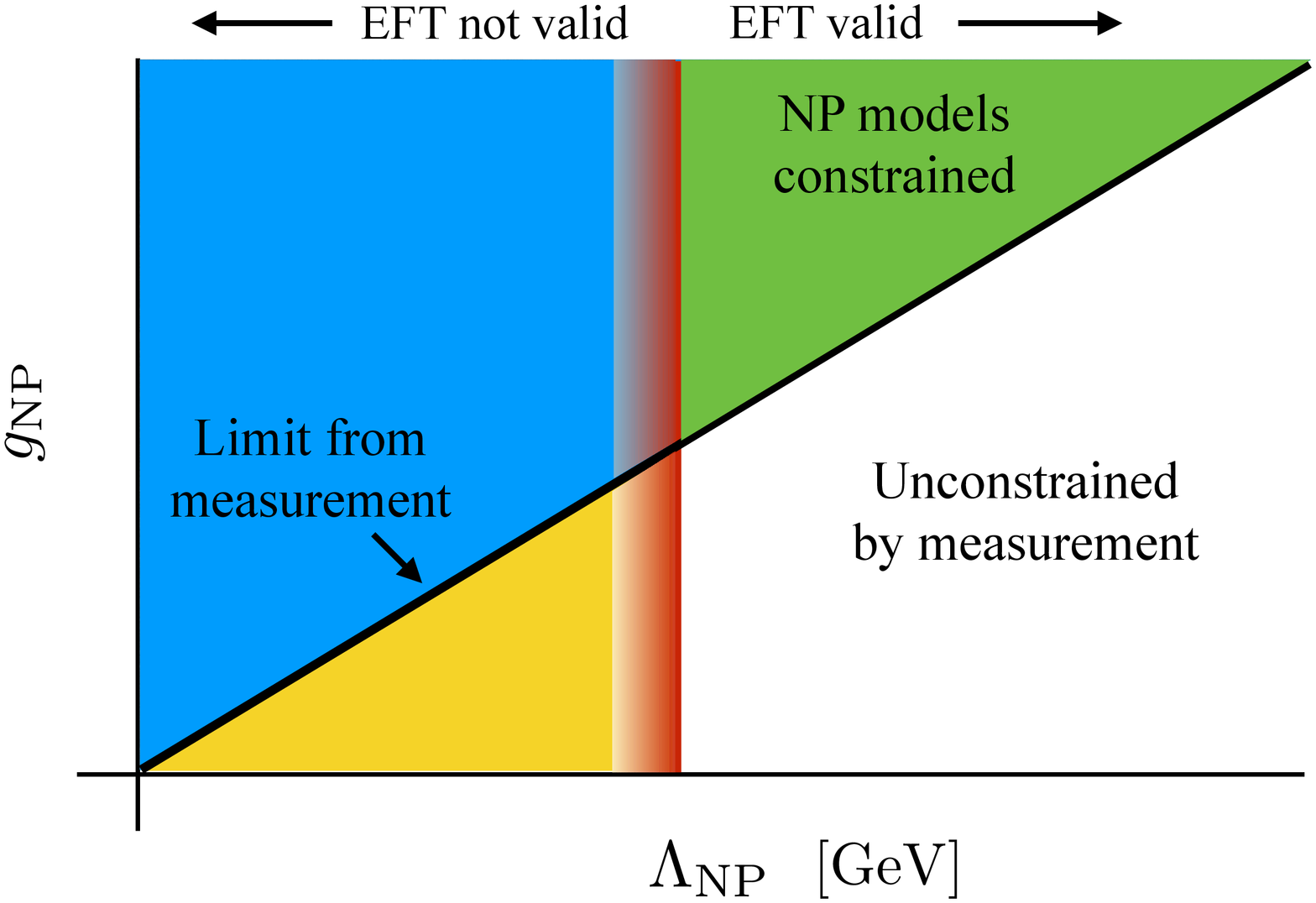}
      \vspace{-0.4cm} }
    \caption{New Physics interpretation of constraint on new operators
      $C(\Lambda_{\text{NP}}) \langle{\hat O}_{\text{NP}}\rangle\sim
      (g_{\text{NP}}/\Lambda_{\text{NP}})^2$ (black line). The red
      vertical line indicates the validity cut-off of the effective
      theory. Only the parameter space captured the by green-shaded
      area is constrained using the effective theory approach.}
  \label{fig:eft}
\end{figure}

To ease the discussion of the examples of
Secs.~\ref{sec:dijets}-\ref{sec:Higgsass} we give here a prescription
of how to obtain constraints on new physics models in terms of
effective theories at particle colliders, taking higher-order
corrections and operator running into account:
\begin{cedescription}
\item[The first step] is of course to perform a measurement of the
  differential distributions relevant for the operators and processes at
  hand, {\it e.g.} $m_{jj}, \Delta \phi_{jj}, p_{T,l}, y_b$. An apt
  choice of the observable is crucial for the sensitivity of the
  analysis.
\item[To constrain new physics models] from the measured observables
  we add higher-dimensional operators, {\it e.g.}
  $\mathcal{L}_\text{dim6}$, to the Lagrangian defined at a new
  physics scale $\Lambda_{\text{NP}}$. Differential distributions
  based on a calculation with the full $\mathcal{L} =
  \mathcal{L}_{\text{dim4}} + \mathcal{L}_{\text{dim6}}$ can now be
  compared to the measured differential distributions.
\item[Obtaining and interpreting constraints] on Wilson coefficients
  can be subtle:
  \begin{mydescription}
  \item[1] When calculating the theory prediction of the differential
    distribution we have to make sure that kinematic regions are
    avoided where the effective theory becomes invalid, {\it i.e.} we
    have to ensure that the effective operators are probed at energies
    below $\Lambda_{\text{NP}}$ always. This can be achieved by
    studying the correlation of the measured distribution with the
    invariant mass $m^2_\text{inv}=( \sum_j p_j )^2$ of the external
    incoming or outgoing particles/fields present in the involved
    operator ${\hat O}_i$. We therefore suggest to record
    $m_\text{inv}$ for every event studied, admittedly a task of
    varying complexity depending on the final state and the operators
    of interest. The maximum value of all recorded invariant masses
    $m^{\text{max}}_\text{inv}$ sets the lower cut-off for
    $\Lambda_{\text{NP}}$ (the red horizontal line in
    Fig.~\ref{fig:eft}), {\it i.e.} the lowest possible scale where
    the effective theory is well-defined. Depending on the size of the
    Wilson coefficient, the obtained limit can still be unphysical if
    unitarity is violated at scales lower than $\Lambda_{\text{NP}}$,
    which is an additional constraint that needs to be
    imposed~\cite{hzdim62}; this fact is reflected in
    Fig.~\ref{fig:eft} by a potentially smeared out region of where
    $\Lambda_{\text{NP}}$ needs to be defined.
  \item[2] After having fixed the upper scale where the effective
    theory is defined, it is worth noting that, because
    $m_{\text{inv}}$ is different for each event, each measured (or
    binned) event probes a different combination of operators. Thus,
    for each measured event one has to relate the combination of
    operators at the measurement scale with the set of operators at
    $\Lambda_\text{NP}$ by solving Eq.~\gl{eq:eqrunn}.
  \item[3] After constraining the Wilson coefficients of an effective
    Lagrangian according to steps~1 and~2, it is now possible to give
    an interpretation of the measurement in terms of new physics
    interactions. As the Wilson coefficients are always a combination
    of dimensionless couplings and powers of the new physics scale
    $\Lambda_{\text{NP}}$, {\it e.g.} for a dimension $6$ operator
    $C_i\langle {\hat O}_i\rangle \sim
    g_\text{NP}^2/\Lambda^2_{\text{NP}}$, the constraint in the
    parameter space corresponds to a diagonal in the
    $g_\text{NP}$-$\Lambda_\text{NP}$ plane, see Fig~\ref{fig:eft}. In
    other words, if the new physics scale is low, small couplings can
    be excluded by the measurement, thereby cutting deep into the
    parameter space of extensions of the Standard Model.
\end{mydescription}
\end{cedescription}
Eventually, four sectors in the $g_\text{NP}$-$\Lambda_\text{NP}$
plane can be identified, separated by the measured constraint on the
Wilson coefficient (black line) and the threshold of the validity
range of the effective theory (red line) in Fig.~\ref{fig:eft}. {\it
  Both lines are inferred directly from the measurement}. The first
sector in the upper left corner (blue shaded area) indicates that the
measurement can rule-out small couplings, however this parameter
choice is outside the validity range of the effective theory
description, as is the yellow-shaded area (we could imagine a New
Physics model with a resonance with smaller mass than
$\Lambda_{\text{NP}}$ to be in the BSM spectrum). The two sectors on
the right from the red line are within the validity range of the
effective theory but only large couplings can be ruled out
(green-shaded area). A large part of the parameter space is not
constrained by the measurement (white-shaded area).

We note that our ignorance of physics at scales higher than the
kinematic LHC cut-off for a given integrated luminosity needs to be
strict in this picture. If we specify a model whose spectrum is
resolved we can always define an effective theory at scales lower than
the lowest new particle mass, but if this mass scale is resolved by
the LHC, the only theoretically correct way to constrain models is to
include the full model dependence on the propagating degrees of
freedom. While the numerical effects can be small depending on the
model, their full inclusion is well possible given the
state-of-the-art of current Monte Carlo event generators.


\section{Dijets and contact interactions at the LHC}
\label{sec:dijets}
Let us come back to the contact interaction model introduced in
Sec.~\ref{sec:eff}. To make our discussion transparent, we use these
results for all contributing quark flavour-changing partonic
subprocesses (and neglect the factor $G_F/\sqrt{2}$ in the operator
definitions). We define the new physics scale and the resulting EFT at
(i) $\Lambda_{\text{NP}}=14~\text{TeV}$, outside the kinematic LHC
coverage of the run 2 start-up energy $\sqrt{s}=13~\text{TeV}$ and
(ii) at the maximum energy of a low statistics phase during run 2
following Sec.~\ref{sec:wilson} in a toy MC analysis. To take into
account the operator mixing and to reflect the energy dependence of
the Wilson coefficients when probed at different centre-of-mass
energies $\sqrt{\hat s}$, we can solve the RGE resulting from
Eqs.~\gl{eq:eqrunn} and \gl{eq:4fermgam} and evaluate the effective
Lagrangian at a specific energy scale on an event-by-event
basis. Setting the correct scale at which we evaluate $\{C_i(\mu)\}$
involves some freedom, similar to choosing an appropriate scale, at
which we evaluate the running of $\alpha_s$ in SM-like simulations of
hadron collider processes. In this particular case we choose
$\mu=\sqrt{\hat{s}}$, which is also chosen to be the relevant scale
for parton densities and the running of the strong coupling.

\begin{figure}[!t]
  \centering
  \includegraphics[width=0.48\textwidth]{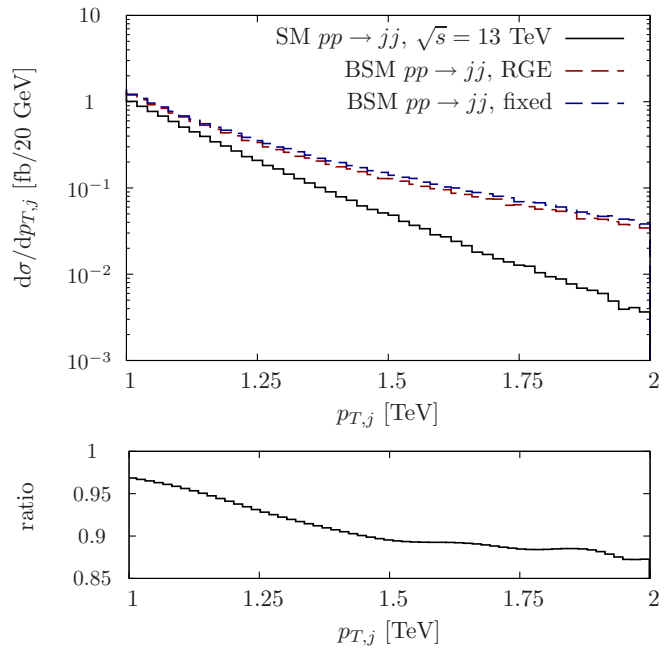}
  \caption{\label{fig:jets} Transverse momentum distribution of dijet
    events at the LHC with $\sqrt{s}=13~\text{TeV}$. We show the SM
    and two scenarios including the effective operators of
    Sec.~\ref{sec:eff}. Scenario 1 (2) refers to a choice of the
    Wilson coefficient of $C_1=C_2=10$. ``fixed'' refers to the
    non-RGE improved distributions and ``RGE'' refers to distributions
    obtained by fixing the effective Lagrangian at
    $\Lambda=14~\text{TeV}$ and using the RGEs to consistently resum
    QCD effects to the measurement scale $\sqrt{\hat s}$. The ratio
    panel gives the differential impact of including the RGE running,
    displaying the ratio of ``fixed'' and ``RGE''.}
\end{figure}

In Fig.~\ref{fig:jets} we display the differential impact of taking
into account the RGE-improved separation of
$\Lambda_{\text{NP}}=14~\text{TeV}$ from the scale at which the
effective Lagrangian is probed as a function of the jets' transverse
momentum $p_{T,j}$.\footnote{These results have been obtained with a
  modified version of MadEvent/MadGraph v5~\cite{Alwall:2011uj},
  inputting a {\sc{Ufo}}~\cite{ufo} model file generated with
  {\sc{FeynRules}}~\cite{feynrules}. We select jets in $|\eta_j|\leq
  2.5$ using the Monte Carlo's default settings. The toy model could
  be thought of in terms of an already constrained very massive $W'$
  boson. We have checked that an analogous $Z'$ model leads to similar
  results.}

Generally the absolute effects dominated over the RGE improved event
simulation as becomes obvious from the logarithmic plot in
Fig.~\ref{fig:jets}. The induced relative difference turns out to be
of order ${O}(10\%)$ in this particular example. Depending on
the size of the data sample and the systematic uncertainty this could
in principle be the level at which the LHC will be able to probe
jet distributions at large luminosities during run 2. 

\begin{figure*}[!t]
  \centering
  \subfigure[]{\includegraphics[width=0.48\linewidth]{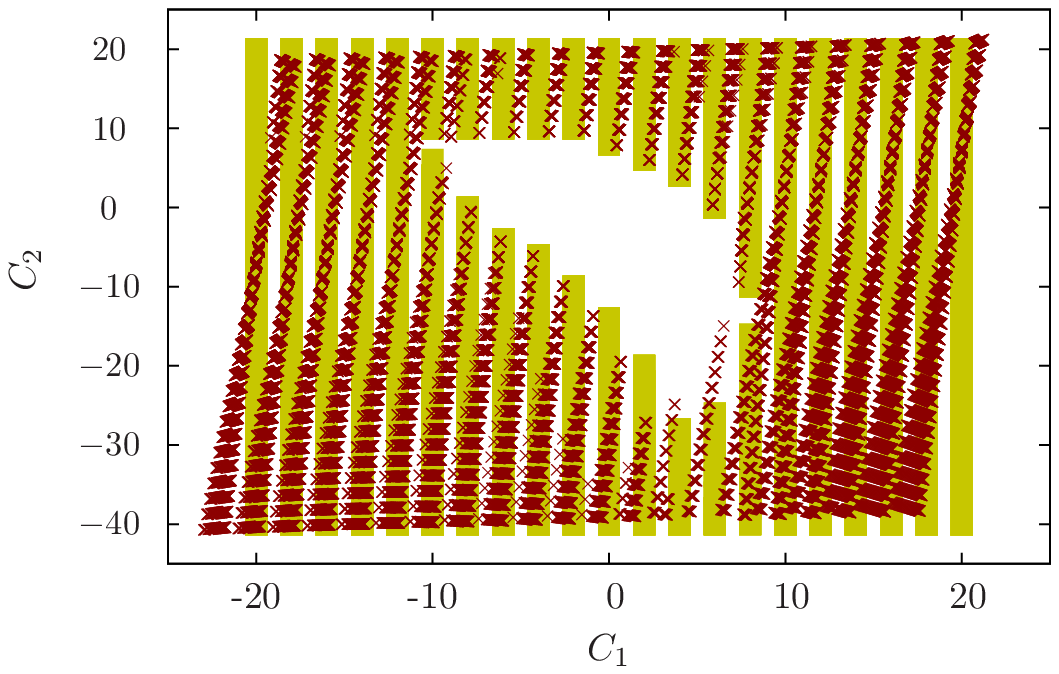}
    \label{fig:jetscattera}}
  \hfill
  \subfigure[]{\includegraphics[width=0.48\linewidth]{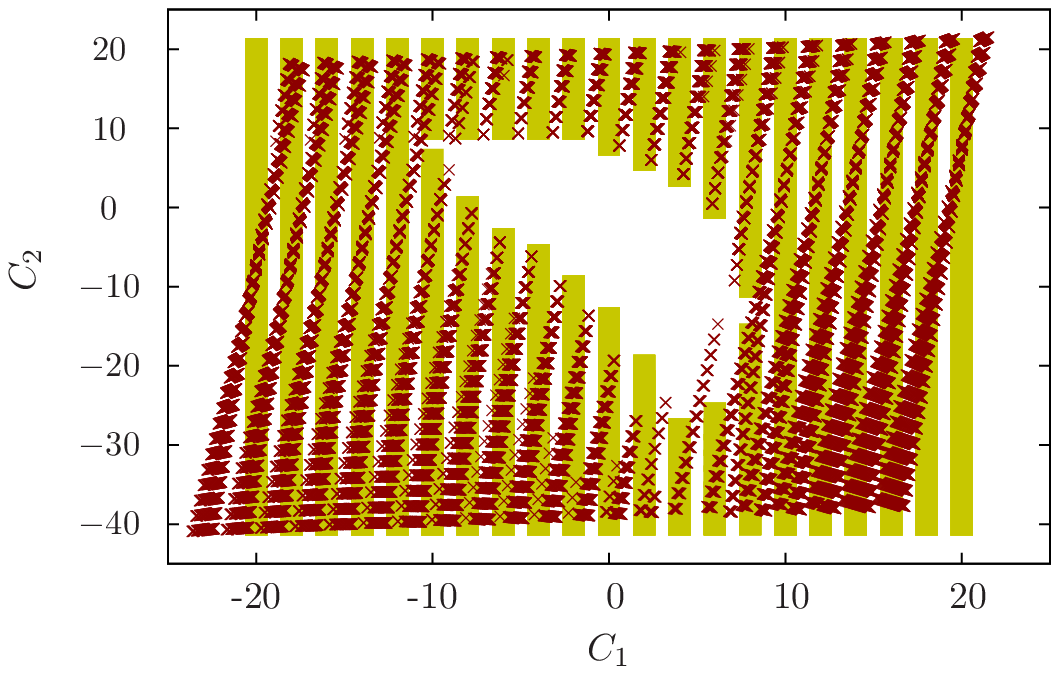}
    \label{fig:jetscatterb}}
  \caption{\label{fig:jetsscatter} Result of the limit setting
    analysis as detailed in the text. Excluded regions are indicated
    by boxes (not including RGE running) and crosses (including the
    RGE evolution to a common scale set by the maximum scale probed in
    a toy Monte Carlo analysis of a sample of ${\cal{L}}=1/\text{fb}$,
    $\Lambda_{\text{NP}}\simeq 8~\text{TeV}$ (a) and outside the run 2
    energy coverage $\Lambda_{\text{NP}}=14~\text{TeV}$ (b)). To allow
    for direct a comparison we rescale the Wilson coefficients by
    $[14~\text{TeV}/\max m_{\text{inv}}]^2$ for (a).}
\end{figure*}

Obviously, for our choice of $\Lambda_\text{NP}$, the impact of RGE
effects are not very large and will not account for the dominant
uncertainties on non-standard interactions at the beginning of run 2
(see Refs.~\cite{dijetNLO,dijetmeas} for a discussion of systematic
uncertainties of jet measurements at the LHC). Given the 10\% relative
impact of a theoretically clean separation of new physics and
measurement scale as demonstrated in Fig.~\ref{fig:jets}, we can turn
the argument around to validate the practitioner's approach of setting
limits on the presence of the new operators without taking into
account the running of RGEs, since their numerical impact is not too
large.

The latter point is demonstrated in Fig.~\ref{fig:jetsscatter}. There
we show a scan of the jet $p_T$ distribution in a toy analysis to set
constraints for new physics effects. Neglecting intricate and
sophisticated experimental techniques to set limits we consider a
parameter point in the $(C_1,C_2)$ as constrained when a bin in the
differential distribution depart from the SM hypothesis by $3\sigma$
at ${\cal{L}}=1/\text{fb}$. We thereby constrain the ``fixed''
distribution of Fig.~\ref{fig:jets} at a certain scale $\mu$; this
yields the yellow box exclusion contour as indicated in
Fig.~\ref{fig:jetsscatter}. The overlayed contour indicated by the
crosses shows how the former contour will be modified if we solve the
RGEs upon inputting the differential measurement. While the overall
modifications can be quite significant, the relative shape between the
two choices of $\Lambda_{\text{NP}}$ is small. Since dijet production
has a large cross section we start to explore the tail of the
distribution very early on during run 2. 


\section{Applications to Higgs Phenomenology}
\subsection{Impact of operator running: Higgs+jet searches}
\label{sec:Higgsjet}
As a first application to Higgs physics and to get an idea of the
typical size of the RGE effects for Higgs phenomenology, we discuss
the impact of operator running on Higgs+jet
production~\cite{uli,schlaffer_spannowsky}. Higgs+jet production is
highly relevant for $H\to \text{invisible}$~\cite{invis} and the
measurement of Higgs couplings in the SM and
beyond~\cite{schlaffer_spannowsky}. While the former scenarios involve
new degrees of freedom at low energy scales, it can be expected that
``genuine'' modifications of Higgs physics result from new dynamics at
scales much higher than the electroweak scale. In fact, if we
interpret the Higgs boson as a pseudo Nambu-Goldstone boson
following~\cite{dim6g}, the new physics scale can easily be pushed to
the multi-TeV regime or even beyond the kinematic LHC coverage if we
admit some degree of fine-tuning. Strong interactions-induced
deviations from the SM Higgs phenomenology will be associated with new
resonant phenomena at the compositeness scale in these scenarios. In
the following we will again assume that those states are outside the
direct sensitivity range of the LHC by defining
$\Lambda_{\text{NP}}=14~\text{TeV}$.

The $pp\to H+\text{jet}$ cross section receives modifications from a
modified Yukawa and effective $ggH$
sector~\cite{uli,schlaffer_spannowsky}. To keep our discussion
transparent at this stage we only focus on the latter operator in the
following ({\it i.e.} we choose like Yukawa interaction
$m_t=y_tv/\sqrt{2}$); to leading logarithmic approximation the two
effective operator contributions are decoupled and the effective $ggH$
sector
\begin{equation}
  \label{eq:hjetop}
  {\hat O}_{G} = {g_s^2\over 2 \Lambda_{\text{NP}}^2} \hat H^\dagger
  \hat H \hat G^{a}_{\mu\nu}\hat G^{a\, \mu\nu}
\end{equation}
gives rise to a form-invariant class of new interactions under RGEs
(we will study the impact of running-induced operator mixing for the
more interesting case of associated Higgs production in the subsequent
section).

The anomalous dimension has been presented in~\cite{trott1}
\begin{equation}
  \label{eq:dimg}
  \gamma_G={1\over 16\pi^2} \left(  -{3\over 2} g'^2 - {9\over 2} g^2 +
    12 \lambda + 6 y_t^2 \right)\,,
\end{equation}
where the authors have used the background field method (note that we
assume a dominant top quark contribution to the Higgs wave function
renormalisation in the following). $\lambda$ denotes the Higgs
self-coupling $V(H^\dagger H)\supset \lambda (H^\dagger H)^2$.

\begin{figure}[!t]
  \centering
  \includegraphics[width=0.48\textwidth]{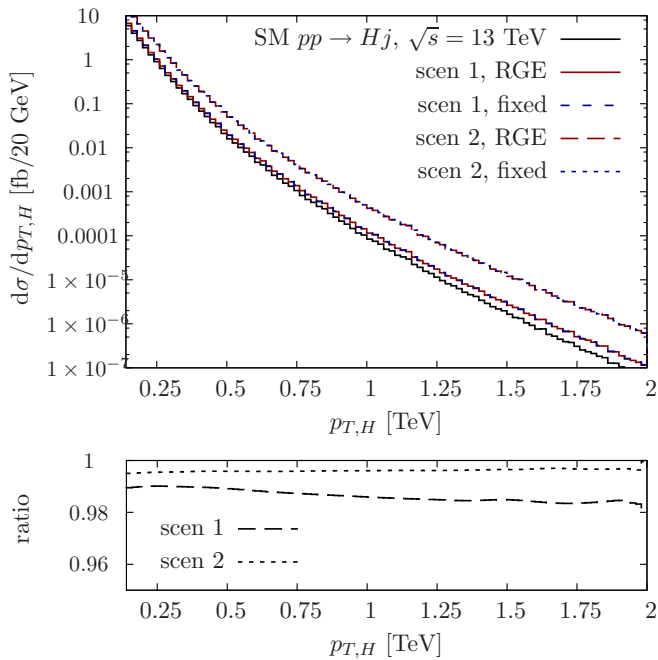}
  \caption{\label{fig:hjdist} Transverse momentum of Higgs bosons
    produced in $pp\to H+\text{jet}$ production for two choices of the
    Wilson coefficients and $\Lambda_{\text{NP}}=14~\text{TeV}$ as detailed
    in the text. The lower panel shows the differential impact of the
    RGE running analogous to Fig.~\ref{fig:jets}.}
\end{figure}

We have validated this result against an independent calculation in
general $R_\xi$ gauge~\cite{Fujikawa:1972fe} using the
{\sc{FeynRules}}~\cite{feynrules} and
{\sc{FeynArts/FormCalc}}~\cite{formcalc} packages. Note that due to
the combination of couplings and gluon field strength tensors in
Eq.~\gl{eq:hjetop}, the anomalous dimension has no dependence on the
strong coupling. This is obvious in the background field
method~\cite{trott1} but non-trivial in $R_\xi$ gauge. To obtain the
result of Eq.~\gl{eq:dimg} we perform a $\overline{\text{MS}}$
renormalisation of the Higgs- and gluon wave functions, as well as of
the strong coupling $g_s$.

Analogous to our discussion in Sec.~\ref{sec:dijets} we show the
impact of the running for two scenarios that correspond to two choices
of Wilson coefficients
\begin{alignat}{5}
  \text{scenario 1:}\quad & C_g= 10\,, \\
  \text{scenario 2:}\quad & C_g= 100\,,
\end{alignat}
for $\Lambda_{\text{NP}}=14~\text{TeV}$, and comparing the differential
impact of the operator running in Fig.~\ref{fig:hjdist}.\footnote{We
  use a purpose-built implementation of $pp\to H+\text{jet}$ based on
  the {\sc{vbfnlo}}~\cite{vbfnlo} framework that includes the full
  numerical solution of the RGE running resulting from
  Eq.~\gl{eq:dimg}. All relevant scales are chosen to be
  $\mu={p_{T,H}}+m_H$.} As it becomes obvious from Fig.~\ref{fig:hjdist}
the RGE effects for $H+\text{jet}$ production are at the 1\% level and
therefore completely negligible in light of expected
theoretical uncertainties in this channel~\cite{hjettheo}. Hence, the
standard limit setting approach is sufficiently adequate.

\begin{figure}[!b]
  \centering
  \includegraphics[width=0.48\textwidth]{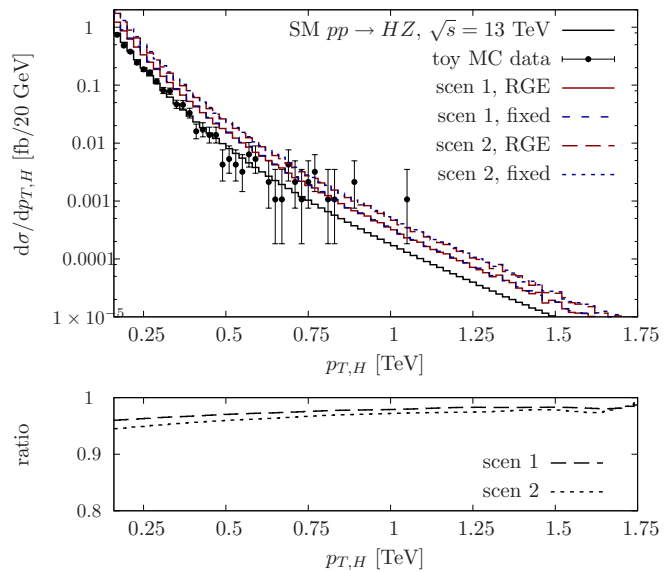}
  \caption{\label{fig:hzpt} Transverse momentum distribution of the
    Higgs boson in the high $p_T$ regime relevant to boosted
    analyses~\cite{boosted} including a toy Monte Carlo data sample
    (for details see text). We show two scenarios referring to
    different choices of the Wilson coefficients that are mixed under
    the RGE flow.}
\end{figure}

\subsection{Impact of Operator Running and Mixing: Higgs Associated
  Production}
\label{sec:Higgsass}
The importance of operator running and mixing in separating IR effects
at the electroweak scale from fundamental physics at a scale
$\Lambda_{\text{NP}}$ has been discussed in the context of the Higgs
branching ratio to photons and electroweak precision observables
in~\cite{trott1,espinosa}. 

A process that turns out to be seminal for the dimension 6 analysis of
the Higgs sector is associated production $pp\to
HZ$~\cite{hzdim6,hzdim62,isidori,Englert:2013vua}. Associated
production has a relatively large cross section and it will typically
be observed at high momentum transfers in boosted final
states~\cite{boosted}, where we can expect new operator contributions
to be well-pronounced. This fact allows to access a plethora of new
physics scenarios in a direct or indirect way~\cite{Englert:2013vua}.

For the sake of clarity we limit ourselves to quark-induced production
and the closed set of operators under
RGEs~\cite{Jenkins:2013zja,trott1}
\begin{alignat}{5}
  {\hat O}_{W} = & {g^2\over 2 \Lambda_{\text{NP}}^2} \hat H^\dagger
  \hat H
  \hat{W}^a_{\mu\nu}\hat{W}^{a\, \mu\nu} \,,\\
  {\hat O}_{B} = & {g'^2\over 2 \Lambda_{\text{NP}}^2} \hat H^\dagger
  \hat H
  \hat{B}_{\mu\nu} \hat{B}^{\mu\nu} \,,\\
  {\hat O}_{WB} =& {gg'\over \Lambda_{\text{NP}}^2} \hat H^\dagger t^a
  \hat H
  \hat{W}^a_{\mu\nu}\hat{B}^{\mu\nu}\,,
\end{alignat}
where $t^a=\sigma^a/2$ are the generators of $SU(2)_L$, $\hat{W}^a_{\mu\nu},
\hat{B}_{\mu\nu}$ are the weak and hypercharge field strength tensor operators,
respectively, with couplings $g$ and $g'$.

The operators ${\hat O}_W$ and ${\hat O}_B$ renormalise the $W^a$ and
$B$ field strengths, and ${\hat O}_{WB}$ measures the departure from
tree-level custodial isospin (the $\rho$ parameter)
$m_W^2/m_Z^2=\cos^2\theta_w+{\cal{O}}(v^4/\Lambda_{\text{NP}}^2)$. Hence,
we can imagine valid models at intermediate scales, such as composite
pseudo Nambu-Goldstone boson interpretations of the Higgs boson, to
incorporate a hierarchy among the Wilson coefficients as discussed in
{\it e.g.} Ref.~\cite{dim6g}.

Again, the anomalous dimension matrix was computed in
Ref.~\cite{trott1}
\begin{widetext}
\begin{equation}
 \label{eq:gamwb}
 \gamma_{WB} = {1\over 16\pi^2} 
 \left[\begin{matrix}
     {1\over 2}g'^2 - {9\over 2}g^2 + 12\lambda +6y_t^2 & 0 & 3g^2 \\
     0 & -{3 \over 2} g'^2 - {5\over 2}g^2 + 12\lambda +6y_t^2 & g'^2 \\
     2g'^2 & 2 g^2 & -{1\over 2}g'^2 + {9\over 2}g^2 + 4\lambda +6y_t^2
   \end{matrix}\right ]\,, 
\end{equation}
\end{widetext}
where we again assume top-yukawa dominance and $\lambda$ is the Higgs
self-coupling $V(H^\dagger H)\supset \lambda (H^\dagger H)^2$. Again,
we have validated this result against an independent calculation in
general $R_\xi$ gauge analogous to Sec.~\ref{sec:Higgsjet}. In $R_\xi$
gauge cancellations between the coupling and field strength
renormalisation constants are non-trivial in the ${\hat O}_W$ and
${\hat O}_{WB}$ cases to yield the gauge-independent result of
Eq.~\gl{eq:gamwb}.

We study the impact of the RGE running for two scenarios fixing
$\Lambda_{\text{NP}}=14~\text{TeV}$,
\begin{alignat}{5}
  \text{scenario 1:}\quad & C_W, C_B= 0.25  {v^2\over\Lambda_{\text{NP}}^2} \,,~ C_{WB} =
  {C_{W,B}/(8\pi^2)}\,,\\
  \text{scenario 2:}\quad & C_W, C_B\simeq 0.50  {v^2\over\Lambda_{\text{NP}}^2} \,,~ C_{WB} = {C_{W,B}/(
    8\pi^2)}\,.
\end{alignat}
which reflects a UV hierarchy to provide and acceptable $\rho$
parameter following~\cite{dim6g}.

The results are shown in Fig.~\ref{fig:hzpt}.\footnote{We use a
  modified version of~\cite{vbfnlo} that includes the full numerical
  solution of the relevant RGEs. All scales are chosen $\mu=\sqrt{\hat
    s}$.} Again the impact of RGE running can be of the order of
$\lesssim 10\%$ in the boosted cut threshold regime where this process
can be isolated from the background~\cite{boosted} and be used to
constrain new interactions~\cite{Englert:2013vua}. There is a mild
dependence of the RGE corrections on the size of the input Wilson
coefficient and due to the particular slope that results from
Eq.~\gl{eq:dimg} the deviations become relevant at low scales where we
can expect the statistical und systematic uncertainties to become
small compared to the $p_T$ distribution's tail eventually. The ratio
quickly converges to one for scales probing ${\cal{O}}({\text{TeV}})$
scales. Therefore, for large luminosities, the separation effects of
$\mu \ll \Lambda_{\text{NP}}$ might be relevant when we will try to
pin down Higgs coupling properties at the 10\% level.

\begin{figure*}[!t]
  \centering
  \subfigure[]{\includegraphics[width=0.48\linewidth]{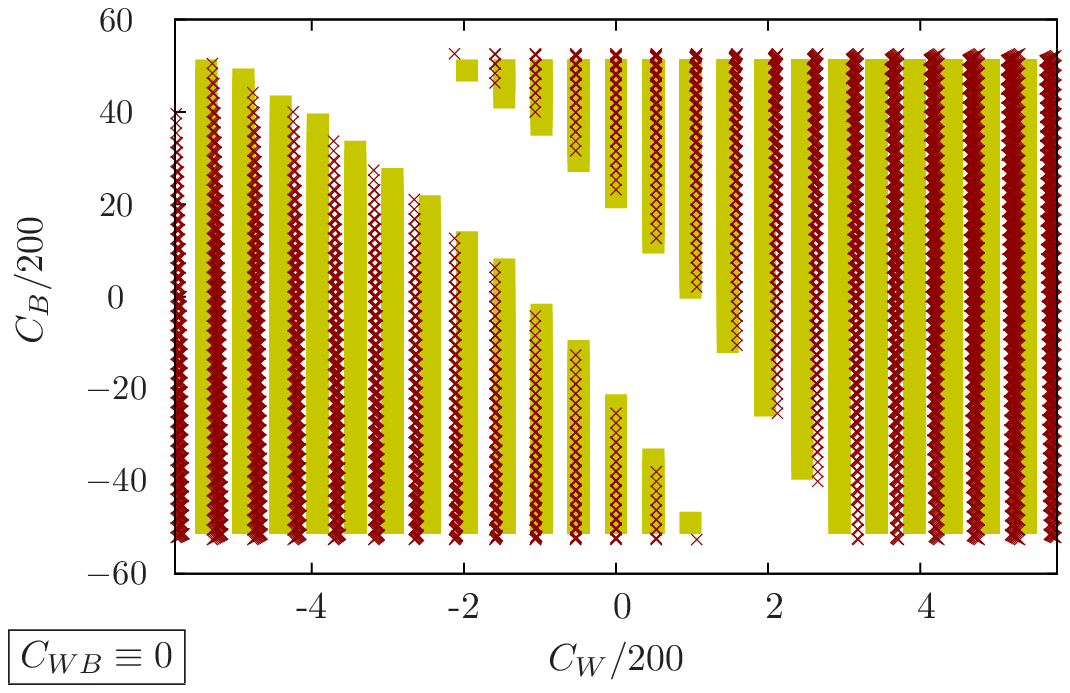}
    \label{fig:hzscattera}}
  \hfill
  \subfigure[]{\includegraphics[width=0.48\linewidth]{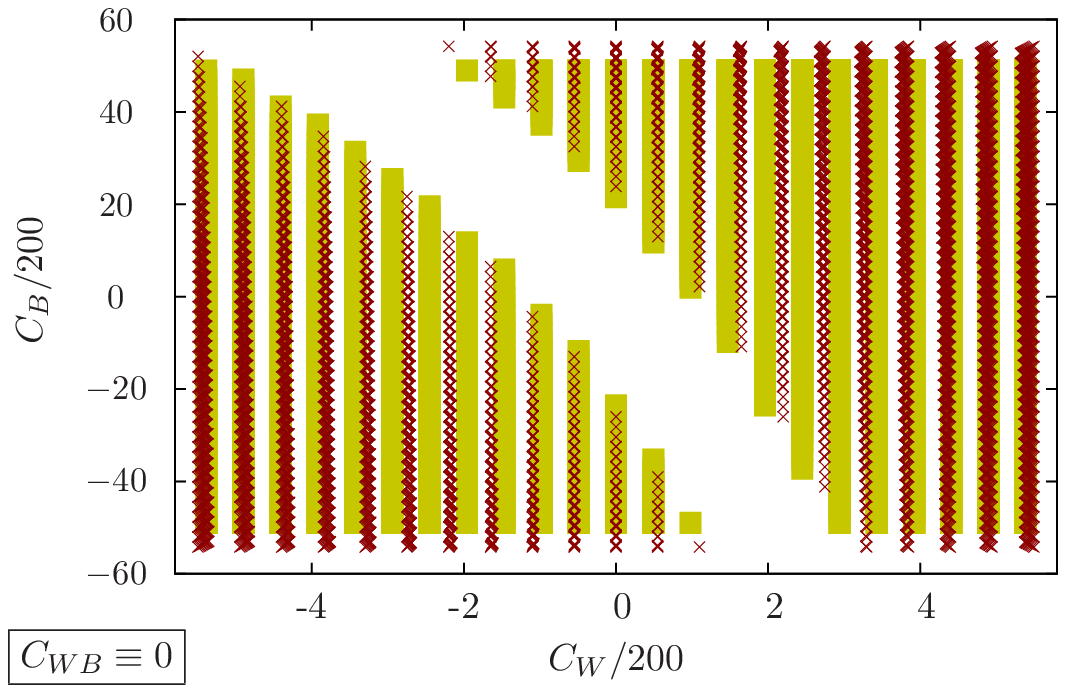}
    \label{fig:hzscatterb}}
  \caption{\label{fig:hzscatter} (a) Scatter plot indicating the
    exclusion contours for $(C_W,C_B,C_{WB}=0)$ from $pp\to HZ$ as
    detailed in the text. We choose $\Lambda_{\text{NP}}\simeq
    2.4~\text{TeV}$, which is the maximum energy scale probed in a toy
    MC experiment with statistics of ${\cal{L}}\simeq 1500/$fb (only
    taking into account branching ratios $Z\to e^+e^-,\mu^+\mu^-$ and
    $H\to b\bar b$) following Sec.~\ref{sec:wilson}. (b) Same as (a)
    but choosing $\Lambda_{\text{NP}}\simeq 14~\text{TeV}$, strictly
    outside the LHC 13 TeV coverage. To allow for direct a comparison
    we rescale the Wilson coefficients by $[14~\text{TeV}/\max
    m_{\text{inv}}]^2$ for (a).}
\end{figure*}

\begin{figure}[!t]
  \centering
  \includegraphics[width=0.48\textwidth]{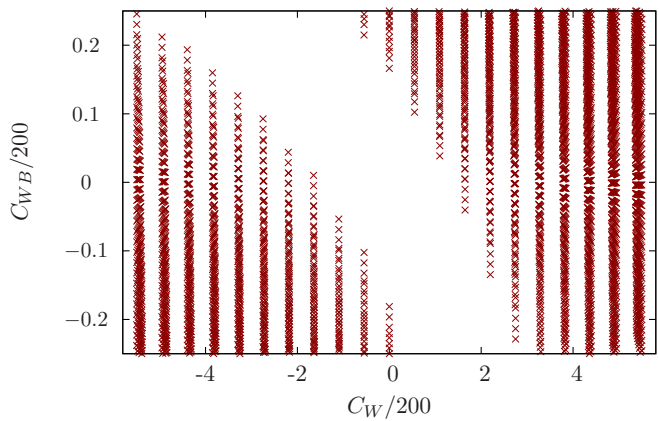}
  \caption{\label{fig:hzscatterinduced}
    Induced $(C_W,C_{WB})$ contour at the scale
    $\Lambda_{\text{NP}}$ that results operator mixing of the scan
    shown in Fig.~\ref{fig:hzscatterb}.}
\end{figure}

\subsection{Interplay between Measurement and Interpretation in Higgs
  Associated Production}
We will use the example of associated production to follow the
description of Sec.~\ref{sec:wilson} more closely. We investigate the
impact on the expected limits numerically with the aim to establish a
connection between measurement, RGE running and the interpretation of
the measurement of in terms of a UV theory.

The first step is to perform a measurement with a given data set, that
determines a maximally resolved scale that probes the operators in the
limit setting exercise. We do this by generating an unweighted SM
event sample of a given luminosity which allows to determine the
maximum $m_{\text{inv}}$ by reconstructing the final state. As we
have discussed in Sec.~\ref{sec:wilson}, $C_i \langle \hat O_i \rangle
\sim g^2_i/\Lambda^2_{\text{NP}}$, and by identifying $\Lambda \simeq
m_\text{inv}^\text{max}$ we assume that the new resonances of the UV
theory are not yet probed with this data set\footnote{Obviously this
  is the most conservative choice, and can be replaced by a
  statistically well-phrased criterion.}. At the same time, however,
we can answer the question of what is the smallest coupling of the UV
theory that we can constrain or exclude in the light of the
measurement (if we deal with a well-defined UV model we can
alternatively rephrase this in terms of a lower bound on the involved
mass scales). This is usually a question of interest: How far into the
parameter space of the UV theory can we cut with this measurement
while being conservative from a new physics perspective.

If the sensitivity is entirely driven by measuring the high $p_T$
phase space region, the impact of operator mixing and running becomes
negligible. In binned log-likelihood hypothesis tests, a significant
amount of sensitivity, however, also stems from lower $p_T$ regions
that are under better systematic and statistic control (we show a toy
MC data sample in Fig.~\ref{fig:hzpt} for comparison).

Obviously, if we choose a cut-off of 14 TeV the impact is more
pronounced.  In most examples we chose 14 TeV, but we stress that this
is a random choice at this stage, which is solely motivated by having
an ad hoc EFT validity over the entire LHC run 2 energy range.

We compare $\Lambda_{\text{NP}}=14~\text{TeV}$ with
$\Lambda_{\text{NP}}= m^{\text{max}}_{\text{inv}}\simeq
2.8~\text{TeV}$ in Fig.~\ref{fig:hzscatterinduced} (for details see
the caption).  Since we only probe a single observable at this stage
we have to make an assumption to reduce the numbers of parameters. We
proceed as outlined in the preceding section to perform a measurement
of $(C_W(\mu)), C_{B}(\mu))$ subject to the boundary condition
$C_{WB}(\mu)=0$. Note that this is merely a choice to obtain an
acceptable $\rho$ parameter at this stage and $C_{WB}$ can be
constrained from other complementary measurements~\cite{spira}
(strictly speaking, the $Z$ mass needs to be input as a boundary
condition to the RGE running).

The difference between choosing $\Lambda_{\text{NP}}$ outside the LHC
coverage and as the maximum available energy is of course that the
larger the ratio of $p_T/\Lambda_{\text{NP}}$ becomes, the more
important the deviation from the standard analysis that does not
include the RGE running becomes.

Even though $C_{WB}=0$ is a boundary condition at the measurement
scale, operator running still induces $C_{WB}\neq 0$ at the UV
scale. To give an estimate of numerical size, we show the induced
exclusion contour in the $(C_W,C_{WB})$ plane for the
$\Lambda_{\text{NP}}=14~\text{TeV}$ in
Fig.~\ref{fig:hzscatterinduced}.


\section{Conclusions}
\label{sec:conc}
Coupling measurements at the 10\% level can be obtained during the LHC
run 2~\cite{review}. This is the level of systematic uncertainty that
can be expected from weak and strong operator running and mixing
effects in the dimension 6 extension of the SM sector and other new
physics scenarios as we have discussed using three instructive
examples. Those particular examples comprehensively discuss the impact
of QCD and electroweak operator mixing and running, especially for a
class of phenomenologically highly relevant operators in the Higgs
sector. As such they stand representative for other (possibly more
complex) processes where we expect our findings to hold qualitatively
as well.  If the RGE-induced effects become of the order of the
expected sensitivity, the resummation effects are relevant in reaching
a consistent interpretation of new physics searches. We stress that
there might well be additional sources of corrections of that size
from additional one-loop effects.

\medskip

A measurement of differential distributions constrains effective
Lagrangians at different energy scales. These measurements can be
consistently combined by using RGEs to evolve results to a
well-defined and separated energy scale. We have outlined such a
programme in Sec.~\ref{sec:wilson}.

For the discussed examples the impact of RGE running are of the order
of $\lesssim 10\%$. If systematic uncertainties in specific channels
turn out to be larger than this figure, our analysis demonstrates that
the standard measurement approach that does not include any RGE
running is perfectly adequate.

In case of systematic uncertainties being under sufficient control, we
encourage the experiments to not only provide a numerical limit on
Wilson coefficients as a result of their measurements, but in addition
the distribution of a characteristic energy scale at which the
operators have been probed as a consequence of our analysis. To give
the measured constraint on the Wilson coefficients an interpretation
in terms of a full UV theory requires to evolve the relevant
coefficients to the theory-intrinsic cut-off scale. However, precisely
this evolution depends on the shape of the differential distribution
of the energy scale at which the operators have been probed during the
measurement.

Our investigation was specific to the LHC run 2 where a vast range of
energy scales will be probed in a fully differential fashion at high
luminosity, but generalises straightforwardly to a future 100 TeV
concept where the discussed phase space effects can be much larger, or
a future linear collider where measurements at the percent-level will
be possible.


\bigskip 

\noindent {\bf{Acknowledgments}} --- We thank C\'eline Degrande for
{\sc{FeynRules}} support and Joachim Brod, Michael Spira and David
Straub for helpful discussions. We thank Uli Nierste for very valuable
comments on the manuscript.

CE thanks Cheikh Anta Diop University Dakar and the attendants and
organisers of the African School of Fundamental Physics 2014, and the
organisers of the ICTP GOAL workshop in S\~ao Paulo for the
hospitality during the time when this work was completed. CE is
supported by the IPPP Associateship programme.

MS thanks the Aspen Center for Physics for hospitality while part
of this work was completed. This work was supported in part by the
National Science Foundation under Grant No. PHYS-1066293.


\end{document}